\documentclass[12pt]{iopart}
\usepackage{graphicx}
\usepackage{iopams}
\usepackage{amssymb}
\usepackage{multirow}
\usepackage{xcolor}

\begin{document}
\title{Standard behaviour of $Bi_{2}Sr_{2}CaCu_{2}O_{8+\delta}$ overdoped}
\author{G.A.~Ummarino}
\ead{giovanni.ummarino@polito.it}
\address{Istituto di Ingegneria e Fisica dei Materiali,
Dipartimento di Scienza Applicata e Tecnologia, Politecnico di
Torino, Corso Duca degli Abruzzi 24, 10129 Torino, Italy; National Research Nuclear University MEPhI (Moscow Engineering Physics Institute),
Kashira Hwy 31, Moskva 115409, Russia}
\begin{abstract}
I calculated the critical temperature and superconducting gap in the framework of one band d wave Eliashberg theory with only one free parameter in order to reproduce the experimental data relative to $Bi_{2}Sr_{2}CaCu_{2}O_{8+\delta}$ $(BSCCO)$ in the overdoped regime. The theoretical calculations are in excellent agreement with the experimental data and indicate that cuprates in the overdoped regime are well described by standard d-wave Eliashberg theory with coupling provided by antiferromagnetic spin fluctuations.

\end{abstract}
%
%
\maketitle
\section{INTRODUCTION}
The properties of cuprates, and of $Bi_{2}Sr_{2}CaCu_{2}O_{8+\delta}$ $(BSCCO)$ in particular, depend strongly on their oxygen content \cite{Watanabe1997PRL,Torsello2018PRMat}. This tunability allows one to study the doping dependence of superconductivity, an approach that has been used to investigate the fundamental properties of several unconventional materials \cite{Pavarini2001PRL,Torsello2019PRB,Wang2020PRB}.
For more than thirty years there has been a major debate in the scientific community about which mechanism is responsible for superconductivity in cuprates. However, most of the research has been focused on the underdoped regime, where there is a variety of competing mechanisms that probably do not specifically concern the superconducting state, but that contribute to confuse and hide the true mechanism of superconductivity.
Several studies underlined all the differences between these new high critical temperature superconductors and the old low critical temperature superconductors whose behavior is perfectly explained by BCS theory or its natural generalization: Eliashberg theory.
In this paper, however, I proposed to highlight the commonalities of cuprates with the old superconductors, not regarding the coupling mechanism but the overall theoretical framework.
In a recent paper \cite{tonicavalla} the authors experimentally investigate the behavior of $BSCCO$ in the overdoped regime by measuring the critical temperature ($T_{c}$), the superconducting gap value ($\Delta_{0}$), the electron boson coupling constant ($\lambda_{Z}$) and the representative energy ($\Omega_{0}$) of the mechanism responsible for superconductivity. Such an experimental study is very useful for theoreticians trying to clarify the mechanism responsible for superconductivity in these materials. Moreover, in the overdoped regime there are superconductors with very high critical temperatures (in our case $T_{c}\leq 91$ K) which make the study interesting in itself.
Always in this paper \cite{tonicavalla} the authors show angle-resolved photoemission spectroscopy studies of $BSCCO$, in the overdoped region up to the to non-superconducting phase. They find that the coupling strength, $\lambda_{Z}$, in the antinodal region of the Fermi surface weakens with doping and at the critical value $\lambda_{Zc}\simeq 1.3$ the superconductivity disappears. This is the evidence that in the overdoped regime, superconductivity
is determined primarily by the coupling strength probably connected with antiferromagnetic spin-fluctuations.
I have tried to reproduce in a comprehensive model these very interesting experimental data.

In Fig. 1 are shown the experimental data present in the paper relative to different values of doping $p$ (the doping away from the half-filling) where the doping is expressed as $p=2A_{FS}-1$ and $A_{FS}$ is area enclosed by the Fermi contour \cite{tonicavalla} that are used as input parameters in the Eliashberg equations. It is possible to see that both quantities: the electron boson coupling constant ($\lambda_{Z}$) and the representative energy ($\Omega_{0}$) decrease with the increase of doping. For comparison with $\lambda_{Z}$, the calculated coupling $\lambda_{\Delta}$ in the gap channel, as it will be explained after, is also shown in the figure.
\label{intro}
%
\section{MODEL}
\label{sec:model}
I calculated the experimental critical temperatures and the superconductive gaps shown in figure 2 by solving the one band d-wave Eliashberg equations \cite{ummarinorev,Dwave1,Dwave2,Dwave3,Dwave4,Dwave5,Dwave6,Dwave7}. In this case two coupled equations for the gap $\Delta(i\omega_{n})$ and renormalization functions $Z(i\omega_{n})$ have to be solved ($\omega_{n}$ denotes the Matsubara frequencies). The d-wave one-band Eliashberg equations
in the imaginary axis representation are:

\begin{eqnarray}
\omega_{n}Z(\omega_{n},\phi)=\omega_{n}+\pi T
\sum_{m}\int_{0}^{2\pi}\frac{d\phi'}{2\pi}\Lambda(\omega_{n},\omega_{m},\phi,\phi')N_{Z}(\omega_{m},\phi')
\end{eqnarray}
\begin{eqnarray}
&&Z(\omega_{n},\phi)\Delta(\omega_{n},\phi)=\pi T
\sum_{m}\int_{0}^{2\pi}\frac{d\phi'}{2\pi}[\Lambda(\omega_{n},\omega_{m},\phi,\phi')-\mu^{*}(\phi,\phi')
\big]\times\nonumber\\
&&\times\Theta(\omega_{c}-|\omega_{m}|)N_{\Delta}(\omega_{m},\phi')
\label{eq:EE2}
\end{eqnarray}
where $\Theta(\omega_{c}-\omega_{m})$ is the Heaviside function, $\omega_{c}$ is a
cut-off energy and
\begin{eqnarray}
\Lambda(\omega_{n},\omega_{m},\phi,\phi')=2\int_{0}^{+\infty}\Omega d\Omega
\alpha^{2}F(\Omega,\phi,\phi')/[(\omega_{n}-\omega_{m})^{2}+\Omega^{2}]
\end{eqnarray}
\begin{eqnarray}
N_{Z}(\omega_{m},\phi)=
\frac{\omega_{m}}{\sqrt{\omega^{2}_{m}+\Delta(\omega_{m},\phi)^{2}}}
\end{eqnarray}
\begin{eqnarray}
N_{\Delta}(\omega_{m},\phi)=
\frac{\Delta(\omega_{m},\phi)}{\sqrt{\omega^{2}_{m}+\Delta(\omega_{m},\phi)^{2}}}
\end{eqnarray}
I assume \cite{ummarinorev,Dwave1,Dwave2,Dwave3,Dwave4,Dwave5,Dwave6,Dwave7} that the electron boson spectral
function $\alpha^{2}(\Omega)F(\Omega,\phi,\phi')$ and the Coulomb
pseudopotential $\mu^{*}(\phi,\phi')$ at the lowest order to contain
separated s and d-wave contributions,
\begin{equation}
\alpha^{2}F(\Omega,\phi,\phi')=\lambda_{s}\alpha^{2}F_{s}(\Omega)
+\lambda_{d}\alpha^{2}F_{d}(\Omega)\sqrt{2}cos(2\phi)\sqrt{2}cos(2\phi')
\end{equation}
\begin{equation}
\mu^{*}(\phi,\phi')=\mu^{*}_{s}
+\mu^{*}_{d}\sqrt{2}cos(2\phi)\sqrt{2}cos(2\phi')
\end{equation}
as well as the self energy functions:
\begin{equation}
Z(\omega_{n},\phi)=Z_{s}(\omega_{n})+Z_{d}(\omega_{n})cos(2\phi)
\end{equation}
\begin{equation}
\Delta(\omega_{n},\phi)=\Delta_{s}(\omega_{n})+\Delta_{d}(\omega_{n})cos(2\phi)
\end{equation}
The spectral functions $\alpha^{2}F_{s,d}(\Omega)$ are normalized in the way that $2\int_{0}^{+\infty}\frac{\alpha^{2}F_{s,d}(\Omega)}{\Omega}d\Omega=1$
and, of course, in this model $\lambda_{Z}=\lambda_{s}$ and $\lambda_{\Delta}=\lambda_{d}$ because in this case
I search for solutions of the Eliashberg equations a
pure d-wave form, as indicated by the experimental data, for the gap function $\Delta(\omega,\phi')=\Delta_{d}(\omega)cos(2\phi)$ (the $s$ component is zero and this happens for example when \cite{varelo} $\mu^{*}_{s}>>\mu^{*}_{d}$). In the more general case $\lambda_{\Delta}$ has $d$ and $s$ components.
The renormalization function $Z(\omega,\phi')=Z_{s}(\omega)$ has just the $s$ component because.
the equation for $Z_{d}(\omega)$ is a homogeneous integral
equation whose only solution in the weak-coupling regime is
$Z_{d}(\omega)=0$ \cite{zetad}. For simplicity we also assume that
$\alpha^{2}F_{s}(\Omega)=\alpha^{2}F_{d}(\Omega)$ and that the
spectral functions is the difference of two Lorentzian, i.e.
$\alpha^{2}F_{s,d}(\Omega)=C[L(\Omega+\Omega_{0},\Upsilon)-L(\Omega-\Omega_{0},\Upsilon)]$
where $L(\Omega\pm\Omega_{0},\Upsilon))=[(\Omega\pm\Omega_{0})^{2}+(\Upsilon)^{2}]^{-1}$,
$C$ is the normalization constant necessary to obtain the proper
values of $\lambda_{s}$, $\Omega_{0}$ and $\Upsilon$ are the peak
energies and half-width, respectively. The half-width is
$\Upsilon=\Omega_{0}/2$. This choice of the shape of spectral function is a good approximation of the true spectral function \cite{a2fd}
connected with antiferromagnetic spin fluctuations. The same thing also happens in the case of iron pnictides \cite{ummarinoiron}.
In any case, even making different choices for $\Upsilon $, it can be verified that as $\Upsilon$ increases, the value of $\lambda_{d}$ decreases and there are no large $\lambda_{d}$ variations for reasonable $\Upsilon$ choices. For example if $\Upsilon=\Omega_{0}$ the reduction of $\lambda_{d}$ is of the order of four percent respect to $\Upsilon=\Omega_{0}/2$. The trend as a function of $\lambda_{d}$ doping remains the same, just the coefficients in the function fit ($\lambda_{d}$ versus $\lambda_{s}$) change a bit.
The cut-off energy is $\omega_{c}=1000$ meV and the maximum quasiparticle energy is $\omega_{max}=1100$ meV.
In first approximation we put $\mu^{*}_{d}=0$ (if the $s$ component of the gap is zero the value of $\mu^{*}_{s}$ is irrelevant).

In this model the experimental input parameters are $\Omega_{0}$ and $\lambda_{s}$ while there is just a free parameter: $\lambda_{d}$.
I solve the imaginary axis d-wave Eliashberg
equations for each value of $\Omega_{0}$ and $\lambda_{s}$ and I seek the value of
$\lambda_{d}$ for obtaining the correct critical temperature that is the most reliable experimental data,
more precisely measured than the value of superconductive gap.
After, via Pad\`{e} approximants \cite{Vidberg}, I calculate the low-temperature value
($T=2$ K) of the gap because, in presence of a strong coupling
interaction, the value of $\Delta_{d}(i\omega_{n=0})$ obtained by solving the imaginary-axis
Eliashberg equations can be very different from the value of
$\Delta_{d}$ obtained from the real-axis Eliashberg equations \cite{Ghigo2017scirep,Dwave3}.
This approach to reproduce experimental data has proved to be very efficient and successful for several materials \cite{Torsello2019prb_mat,Torsello2019JOSC}.
\section{RESULTS AND DISCUSSION}
In Fig. 2 the results are shown. From Fig. 2 it is possible see that the critical temperatures are perfectly reproduced and the behaviour of gap with the doping well enough: all in the framework of a very simple model without nothing of "exotic". What may create some perplexity are the large values of coupling constant for doping values close to optimal doping.
Probably the values of the coupling constants are effective values \cite{MigdalBen} because in this model I do not take into account the violation of the Migdal theorem \cite{Migdal1} that almost certainly happens. In fact, it has been shown that using an Eliashberg theory generalized to the case in which the Migdal theorem is not valid, one obtains values of the coupling constants $\lambda_{\Delta}$ and $\lambda_{Z}$ that are much smaller than those used in the standard Eliashberg theory to produce the same critical temperatures \cite{Migdal}.
I find that the link between $\lambda_{d}$ (free parameter, determined by the calculus of critical temperature exactly equal to experimental one) and $\lambda_{s}$ (experimental data, input parameter) (see figure 1) is reproduced very well by the equation
\begin{equation}
\lambda_{d}=2.1881(\lambda_{s}-1.3)^{0.6021}
\end{equation}
Also from this formula it is possible to see that the critical value
of $\lambda_{s}$ is $\lambda_{sc}=1.3$ as I see from experimental data \cite{tonicavalla}. The calculation of the critical temperature from the solution of Eliashbeg equations in this case depends strongly from the values of $\lambda_{d}$ and $\lambda_{s}$ and also small differences in the values of coupling constants produce large variation in the critical temperature. For this reason in the formula of $\lambda_{d}$ there are four decimal digits.
In Fig 3 the values of $\Delta_{d}(i\omega_{n=0})$ as a function of temperature are shown.
\label{sec:results}
\section{CONCLUSIONS}
In this paper I have shown that the experimental data ($T_{c}$ and $\Delta_{0}$) in the overdoped regime for $BSCCO$ can be reproduced by a very simple model: the standard one band d wave Eliashberg equations with coupling provided by antiferromagnetic spin fluctuations. This indicates that the superconducting state has no particular characteristics in the overdoped regime.
The fact that this simple model explains very well the experimental data could be a new stimulus for further theoretical investigations in the underdoped regime that neglects a lot of exotic competing orders present in the normal state that probably are not directly connected with the superconducting phase.
\label{sec:conclusions}
\ackn
The author acknowledges support from the MEPhI Academic Excellence Project (Contract No. 02.a03.21.0005) and dr. D. Torsello for useful suggestions.\\

\newpage
\begin{figure}
\begin{center}
\includegraphics[keepaspectratio, width=\columnwidth]{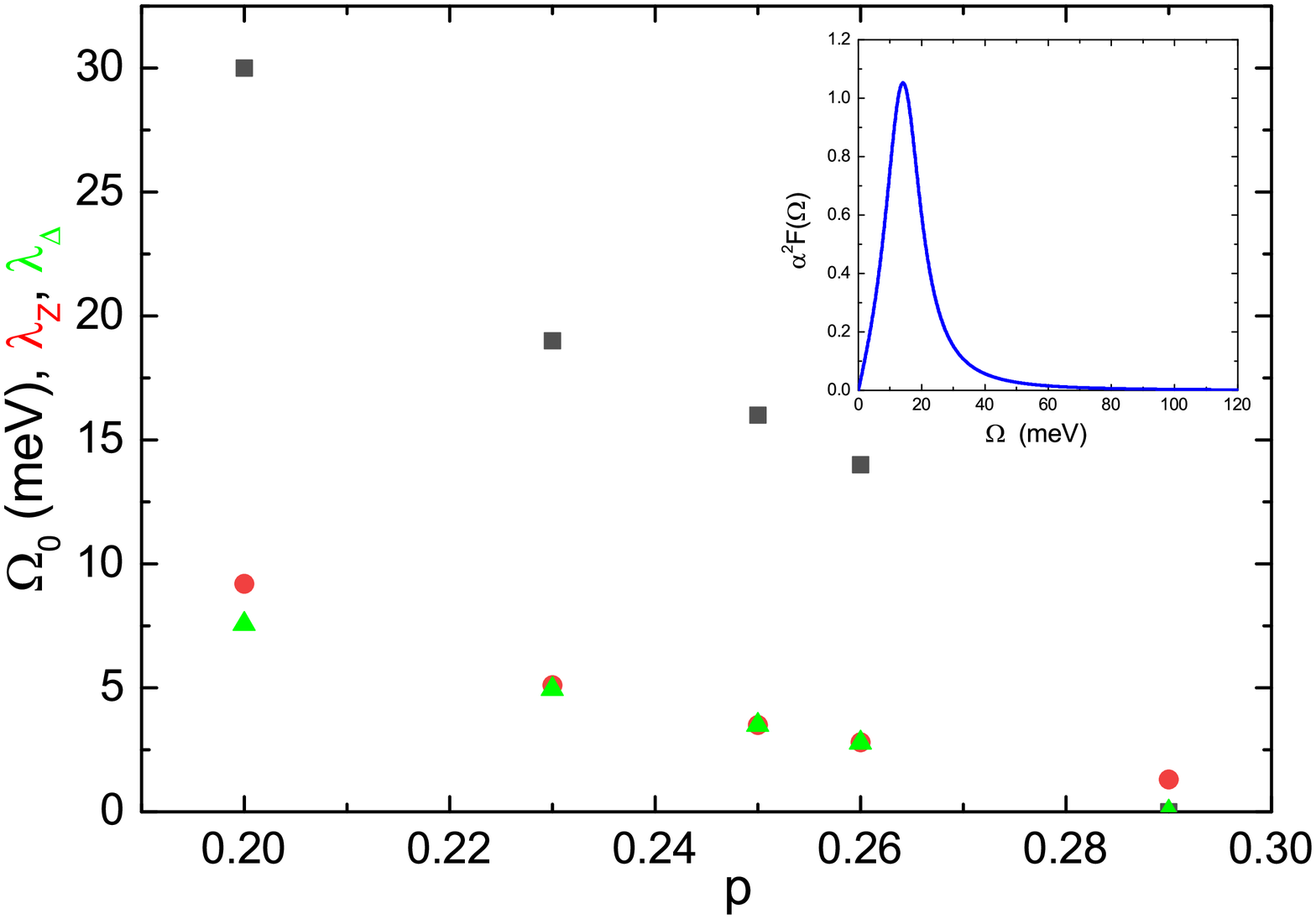}
\vspace{-5mm} \caption{(Color online)
 Experimental input parameters of Eliashberg theory: representative energy $\Omega_{0}$ (black full squares) and electron-boson coupling constant $\lambda_{Z}$ (red full circles) and $\lambda_{\Delta}$ (green full triangles) as a function of doping $p$. $\lambda_{\Delta}$ is the value obtained, via Eliashberg equations, for reproducing exactly the experimental critical temperatures. In the inset the electron-boson spectral function for the last value of doping $p=0.26$ is shown.
 }\label{Figure1}
\end{center}
\end{figure}
\newpage
\begin{figure}
\begin{center}
\includegraphics[keepaspectratio, width=\columnwidth]{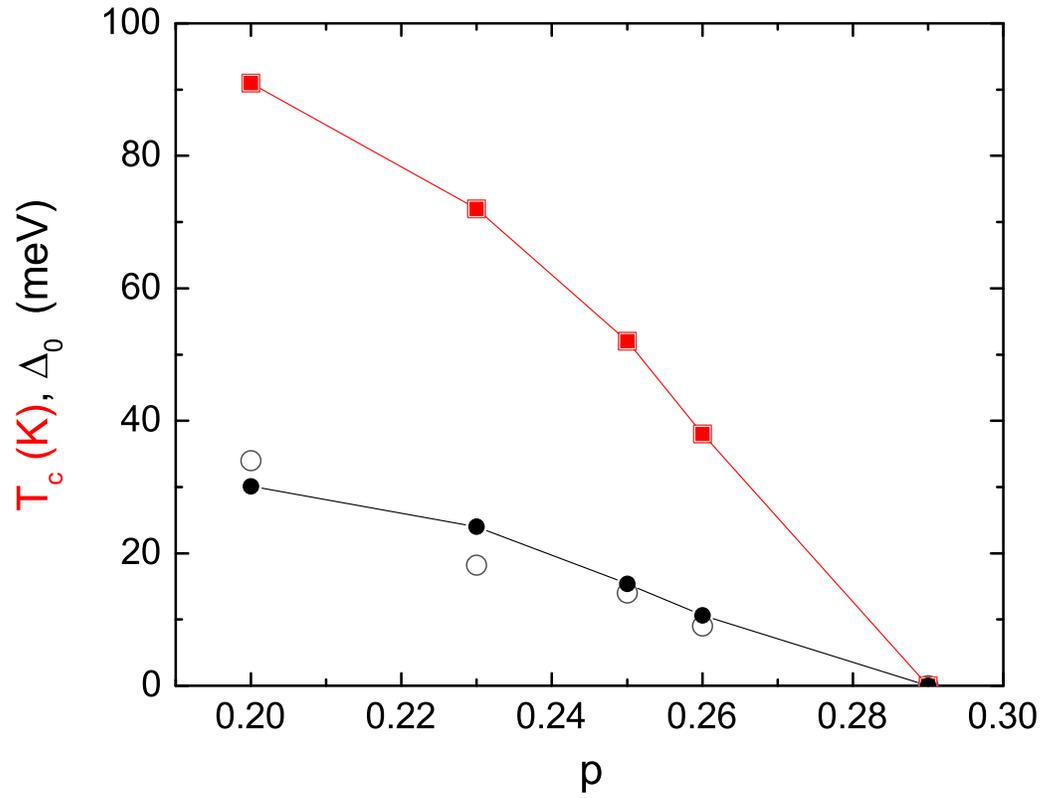}
\vspace{-5mm} \caption{(Color online)
 Calculated values of critical temperature (full red squares) and superconductive gap (full black circles) compared with experimental data (open red squares for the critical temperature and open black circles for the superconductive gap) in function of the doping $p$. The lines are guides for the eye.
 }\label{Figure2}
\end{center}
\end{figure}
\newpage
\begin{figure}
\begin{center}
\includegraphics[keepaspectratio, width=\columnwidth]{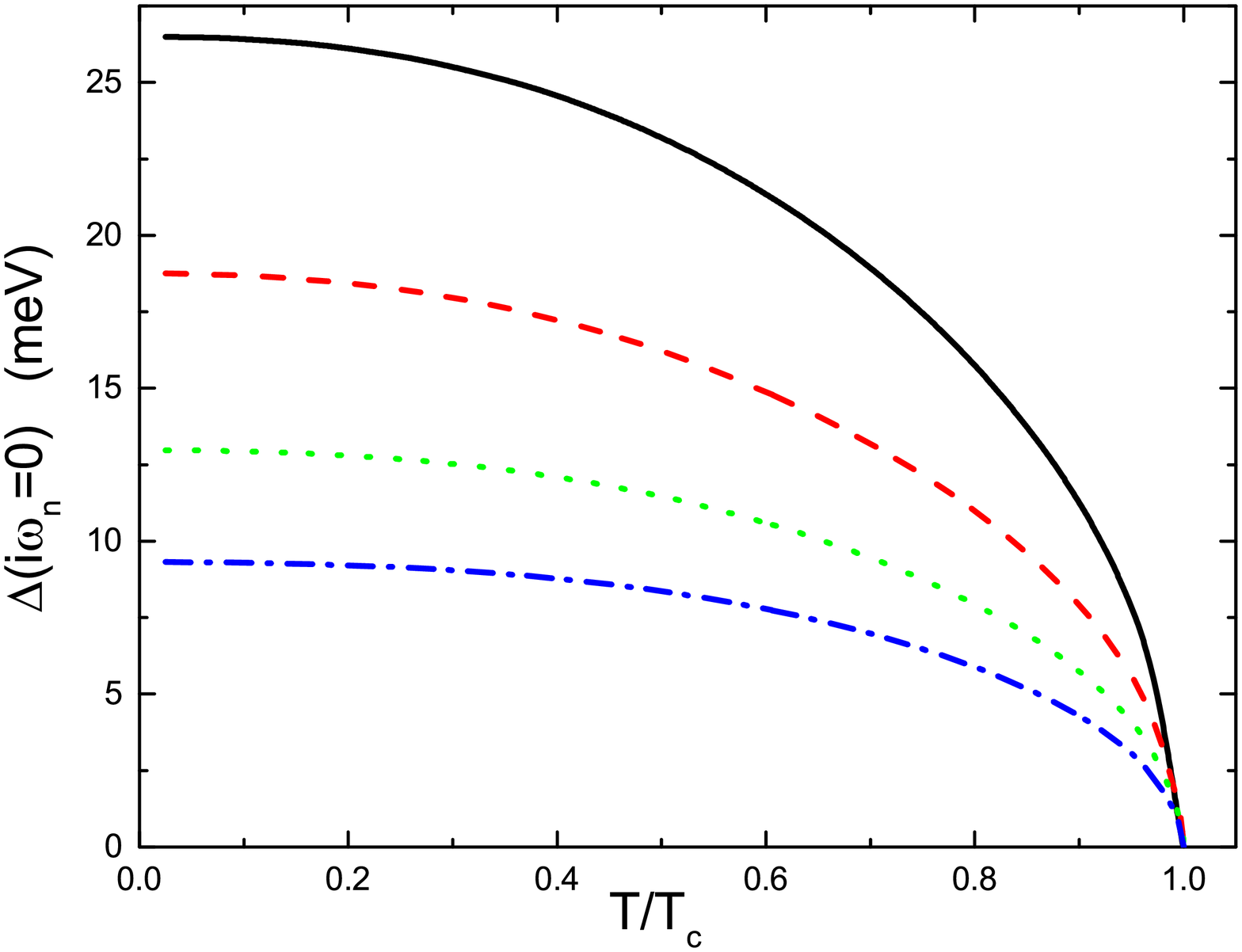}
\vspace{-5mm} \caption{(Color online)
 Calculated values of $\Delta_{d}(i\omega_{n=0})$ as a function of normalized temperature $(T/T_{c})$ for the four cases examinated: $p=0.20$ black solid line, $p=0.23$ red dashed line, $p=0.25$ green dotted line and $p=0.26$ dark blue dashed dotted line.
 }\label{Figure3}
\end{center}
\end{figure}


\begin{thebibliography}{99}
\small{

\bibitem{Watanabe1997PRL} T. Watanabe, T. Fujii, A. and Matsuda, \textit{Phys. Rev. Lett.} \textbf{79}, 2113 (1997).
%
\bibitem{Torsello2018PRMat} Daniele Torsello, Lorenzo Mino, Valentina Bonino, Angelo Agostino, Lorenza Operti, Elisa Borfecchia, Ettore Vittone, Carlo Lamberti, Marco Truccato, \textit{Phys. Rev. Materials} \textbf{2}, 014801 (2018).
 %
\bibitem{Pavarini2001PRL} E. Pavarini, I. Dasgupta, T. Saha-Dasgupta, O. Jepsen, O.K. Andersen,
\textit{Phys. Rev. Lett.} \textbf{87}, 047003 (2001).
%
\bibitem{Torsello2019PRB} D. Torsello, K. Cho, K. R. Joshi, S. Ghimire, G.A. Ummarino, N.M. Nusran, M.A. Tanatar, W.R. Meier, M. Xu, S.L. Bud'ko, P.C. Canfield, G. Ghigo, R. Prozorov, \textit{Phys. Rev. B} \textbf{100}, 094513 (2019).
%
\bibitem{Wang2020PRB} Zhan Wang, Zhang Guang-Ming, Yi-feng Yang, Fu-Chun Zhang,
\textit{Phys. Rev. B} \textbf{102}, 220501 (2020).
%
\bibitem{Ghigo2017scirep} G. Ghigo, G.A. Ummarino, L. Gozzelino, R. Gerbaldo, F. Laviano, D. Torsello, T. Tamegai,
\textit{Sci. Rep.} \textbf{7}, 13029 (2017).
 %
\bibitem{Torsello2019prb_mat} D. Torsello, G.A. Ummarino, L. Gozzelino, T. Tamegai, G. Ghigo,
 \textit{Phys. Rev. B} \textbf{99}, 134518 (2019).
 %
\bibitem{Torsello2019JOSC} Daniele Torsello, Giovanni Alberto Ummarino, Roberto Gerbaldo, Laura Gozzelino, GianLuca Ghigo,
\textit{Journal of Superconductivity and Novel Magnetism} \textbf{33}, 2319 (2020).
\bibitem{tonicavalla} T. Valla, I.K. Drozdov, G.D. Gu, \textit{Nat Commun} \textbf{11}, 569 (2020).
%
\bibitem{ummarinorev} G.A. Ummarino, Eliashberg Theory. In: \textit{Emergent Phenomena in Correlated Matter}, edited by E. Pavarini, E. Koch, and U. Schollw\"{o}ck, Forschungszentrum J\"{u}lich GmbH and Institute for Advanced Simulations, pp.13.1-13.36 (2013) ISBN 978-3-89336-884-6.
%
\bibitem{Dwave1} C.T. Rieck, D. Fay, L. Tewordt, \textit{Phys. Rev. B} \textbf{41}, 7289 (1989).
    %
\bibitem{Dwave2}  G.A.Ummarino and R.S. Gonnelli, \textit{Physica C} \textbf{328}, 189 (1999).
 %
\bibitem{Dwave3}  G.A.Ummarino and R.S. Gonnelli, \textit{Physica C} \textbf{341-348}, 295 (2000).
%
\bibitem{Dwave4} G.A. Ummarino, D. Daghero and R.S. Gonnelli, \textit{Physica C} \textbf{377}, 292 (2002).
%
\bibitem{Dwave5} E. Cappelluti, G.A. Ummarino, \textit{Phys. Rev. B} \textbf{76}, 104522 (2007).
%
\bibitem{Dwave6} F. Jutier, G.A. Ummarino, J.C. Griveau, F. Wastin, E. Colineau, J. Rebizant, N. Magnani, and R. Caciuffo, \textit{Phys. Rev. B} \textbf{77}, 024521 (2008).
    %
\bibitem{Dwave7} G.A. Ummarino, R. Caciuffo, H. Chudo and S. Kambe, \textit{Phys. Rev. B} \textbf{82}, 104510 (2010).
%
\bibitem{varelo} Georgios Varelogiannis, \textit{Solid State Communications} \textbf{107}, 427 (1998).
%
\bibitem{a2fd} Jin Mo Bok, Jong Ju Bae, Han-Yong Choi, Chandra M. Varma, Wentao Zhang,
Junfeng He, Yuxiao Zhang, Li Yu, X. J. Zhou, \textit{Sci. Adv.} \textbf{2}, 1501329 (2016)
%
\bibitem{ummarinoiron} G.A. Ummarino, \textit{Phys. Rev. B} \textbf{83}, 092508 (2011).
%
\bibitem{zetad} K. A. Musaelian, J. Betouras, A. V. Chubukov, and R. Joynt, \textit{Phys. Rev. B} \textbf{53}, 3598 (1996).
%
\bibitem{Vidberg} H. Vidberg and J. Serene, \textit{J. Low Temp. Phys.} \textbf{29} 29, 179 (1977).
%
\bibitem{MigdalBen} P.Benedetti, C. Grimaldi, L. Pietronero and G. Varelogiannis,  \textit{Europhys. Lett} \textbf{28}, 251 (1994).
%
\bibitem{Migdal1} L. Pietronero, S. Strassler, and C. Grimaldi, \textit{Phys. Rev. B} \textbf{52}, 516 (1995);
C. Grimaldi, L. Pietronero, and S. Strassler, \textit{Phys. Rev. B} \textbf{52}, 530 (1995).
%
\bibitem{Migdal} G.A.Ummarino and R.S. Gonnelli, \textit{Phys. Rev. B} \textbf{56}, 14279 (1997).

}\end{thebibliography}
\end{document}